\documentclass[aps,pre,twocolumn,showpacs,preprintnumbers,a4paper]{revtex4}

\usepackage[nointlimits]{amsmath} % use ams extensions
\usepackage[latin1]{inputenc}     % allow for special input symbols (linux encoding)
\usepackage{graphicx}             % for inclusion of external graphics
\usepackage{subfigure}            % store multiple figures in one main environment
\usepackage{color}                % allow for colored text

\newcommand*{\lavg}{\left\langle}

\newcommand*{\ravg}{\right\rangle}

\newcommand*{\vect}[1]{\mathbf{#1}}

\begin{document}

\preprint{LMU-ASC 32/06}
 
\title{Conformations of Confined Biopolymers}

\author{Frederik Wagner$^{1}$, Gianluca Lattanzi$^2$, Erwin Frey$^1$}

\affiliation{$^1$Arnold Sommerfeld Center for Theoretical Physics
  (ASC) and Center for NanoScience (CeNS),
  Ludwig-Maximilians-Universit\"at M\"unchen,
  Theresienstra\ss e 37, D-80333 M\"unchen, Germany \\
  $^2$Department of Medical Biochemistry, Biology and Physics,
  TIRES-Center and INFN, Universit\`a di Bari, Piazza Giulio Cesare
  11, 70124 Bari, Italy}

\date{\today}

\begin{abstract}
  Nanoscale and microscale confinement of biopolymers naturally occurs in cells
  and has been recently achieved in artificial structures designed for
  nanotechnological applications. Here, we present an extensive theoretical
  investigation of the conformations and shape of a biopolymer with varying
  stiffness confined to a narrow channel. Combining scaling arguments,
  analytical calculations, and Monte Carlo simulations, we identify various
  scaling regimes where master curves quantify the functional dependence of the
  polymer conformations on the chain stiffness and strength of confinement.
\end{abstract}

\pacs{87.16.Ac, 36.20.Ey, 82.35.Lr, 87.16.Ka}

% 87.16.-b: Subcellular structure and processes
% 87.16.Ac: Biomolecules, Theory and modeling; computer simulation
% 87.16.Ka: Filaments, microtubules, their networks, and supramolecular assemblies
% 36.20.Ey: Conformation (statistics and dynamics) macromolecules and polymers
% 82.35.Lr: Physical properties of polymers

\maketitle

% General introduction
What is the effect of confinement on the shape of a biopolymer? With recent
advances in visualizing and manipulating macromolecules on ever shrinking length
scales, an answer to this question has gained increasing importance. In the
crowded environment of a cell the conformations of cytoskeletal filaments are
highly constrained by other neighboring macromolecules. 
This confinement largely alters the viscoelastic response of
entangled biopolymer solutions~\cite{ClaessensEtAl2006, HinnerEtAl1998}. There
is growing interest in manufacturing nanostructures such as
nanopores~\cite{Austin2003} and nanochannels~\cite{ReisnerEtAl2005,
TegenfeldtEtAl2004} for investigating and manipulating DNA with improved
technologies aiming towards smaller and smaller structures. Hence an improved
understanding of the effect of confinement on biopolymer conformations has
potential implications for the design of nanoscale devices in biotechnological
applications. Similarly, microfluidic devices have been used to explore
confinement effects on actin filaments and
DNA~\cite{KoesterSteinhauserPfohl2005, BalducciEtAl2006}. What makes the
confinement of biopolymers both a challenging and interesting problem is that
biopolymers, unlike their synthetic counterparts, are generally stiff on a
length scale much larger than their monomer size. The persistence length
$\ell_p$, the scale below which bending energy dominates over thermal
fluctuations, is approximately $50\,\text{nm}$ for DNA~\cite{BouchiatEtAl1999}
and $16\,\mu$m for F-actin~\cite{LeGoffEtAl2002}. Depending on whether the
contour length $L$ is smaller or larger than the persistence length we may
distinguish between stiff and flexible chains.

%fig_1
\begin{figure}[b]
  \centering
  \includegraphics{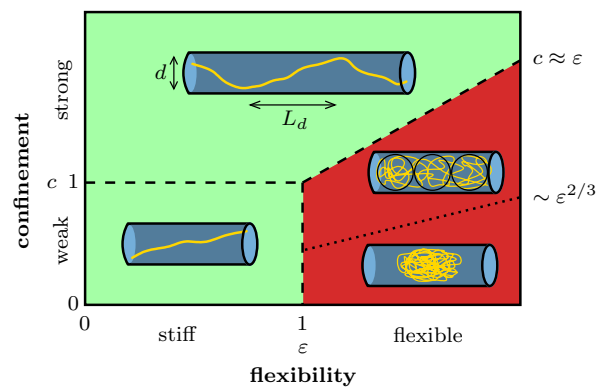}
  \caption{(Color online) Scaling regimes for confined biopolymer conformations as a function
    of polymer flexibility $\varepsilon=L/\ell_p$ and confinement strength
    $c=L/L_d$. In the flexible regime ($\varepsilon \gg 1$) two scaling regimes
    are known [\emph{dark grey} (\emph{red})]: free coil behavior for weak confinement and de
    Gennes scaling for intermediate confinement. In the parameter range most
    relevant for biopolymers [\emph{light gray} (\emph{green}], one has to distinguish between
    weak confinement of stiff polymers and strong confinement for chains of
    arbitrary stiffness.}
  \label{fig:sketch}
\end{figure}
%fig_1

%Scaling scenarios
For cellular systems as well as for nanoscale devices, biopolymers are confined
on length scales comparable with their persistence length $\ell_p$ such that the
polymer's intrinsic bending stiffness plays a decisive role for its
conformations.  For simplicity, consider a cylindrical tube of diameter $d$.
Upon balancing the bending stiffness of a chain with thermal energy,
Odijk~\cite{Odijk1983} has identified a length $L_d$ measuring the typical
distance between successive deflections of the chain, $L_d^3 \sim d^2 \ell_p$;
see Fig.~\ref{fig:sketch}. This suggests to use the \emph{number of collisions}
$c=L/L_d$ per filament length $L$ as a natural dimensionless parameter to
measure the strength of confinement and $\varepsilon = L/\ell_p$ to measure the
\emph{flexibility} of a polymer. The physics in the \emph{strong confinement} regime
($c \gg 1$) is genuinely different from the regime where the radius of gyration
$R_\text{G}$ of a long flexible chain (with $L \gg \ell_p$) becomes comparable
to $d$. In the latter case of \emph{weak confinement} of a flexible chain the
shape of the polymer is distorted due to self-avoidance between distant segments
along the polymer chain. Then, according to de Gennes' blob
picture~\cite{DeGennes1979}, we may represent the conformation of the polymer as
a linear chain of nonpenetrating spheres of radius $d$, where each sphere is
described by Flory's theory. This picture results in the following scaling law
for an extension of the polymer along the tube axis:
\begin{equation}
  R_\parallel \sim L \left(\ell_p / d \right)^{2/3} \, .
\end{equation}
For strong confinement the Odijk length becomes the analogue of the blob size,
below which the polymer may be considered as free. Thus the fraction of
contour length stored in thermal undulations decreases with decreasing tube
dimensions~\cite{Odijk1983}:
\begin{equation}
  (L- R_\parallel)/L \sim L_d / \ell_p \, .
\end{equation}
This scaling law should apply equally well for stiff and flexible chains as long
as the collision parameter $c$ is sufficiently large. For flexible chains,
$\varepsilon \gg 1$, this is the case if the deflection length $L_d$ is less
than the persistence length $\ell_p$. For stiff chains, $\varepsilon \ll 1$, the
deflection length has to become smaller than the total filament length before
there is any stretching. There is an additional regime of weak confinement ($d
\leq L \leq L_d$) where the average orientation of the filament becomes aligned
with the tube axis. These various scaling scenarios are summarized in
Fig.~\ref{fig:sketch}.

%Model
The purpose of this Rapid Communication is to go beyond this qualitative scaling
picture and provide a quantitative study of the conformations of biopolymers in
confined geometry. Our focus is on the parameter range that is most relevant for
cellular systems and nanoscale devices [light gray (green) region in
Fig.~\ref{fig:sketch}],
where self-avoidance effects may safely be neglected. For specificity, we
consider a wormlike chain in a soft harmonic potential of cylindrical symmetry
and strength $\gamma$. Indeed, the harmonic potential is the simplest model to
represent a tubelike confinement and it has the advantage of being amenable to
analytic calculations. Thus the Hamiltonian for the contour $\vect r (s)$
parametrized in terms of the arc length $s$ reads
\begin{equation}
  \mathcal H[\vect r(s)]  = \frac\kappa2\int_0^Lds\left(\frac{\partial^2 \vect
  r(s)}{\partial s^2}\right)^2 + \frac\gamma2\int_0^Lds\,\vect r_\perp^2(s)\,,
\end{equation}
where $\kappa = \ell_p k_\text{B} T$ is the bending stiffness, and $\vect
r_\perp =\left(x, y \right)$ are the components of the contour perpendicular to
the tube axis.

% Analytical results for unconfined chains
A variety of analytical results have been obtained so far for unconfined
wormlike chains.  For instance, the tangent-tangent correlation
functions~\cite{LandauLifschitz1979} and moments of the end-to-end distributions
have been calculated exactly~\cite{SaitoTakahashiYunoki1967,YamakawaFujii1973}.
Further results like the probability distribution function of the end-to-end
distance $\vect R$ have been calculated for stiff chains~\cite{WilhelmFrey1996}
within the weakly bending rod (WBR) approximation, where one considers only
small transverse bending fluctuations with respect to a straight contour
$\vect{r} (s) \approx \left(\vect r_\perp (s), s\right)$. For confined chains
the WBR limit amounts to assume that the filament is aligned with the tube axis,
and may be employed to study the asymptotic regime of strong confinement, $L_d
\ll L$. Then, using the equipartition theorem one obtains for the correlation
function of the transverse undulations in Fourier space:
\begin{equation}  \label{eq:mode_spectrum}
  \lavg x(k) x(-k) \ravg = \frac{k_\text{B}T}{\kappa (k^4 + 4 L_d^{-4})}\,,
\end{equation}
where $L_d = (4 \kappa / \gamma)^{1/4}$. This simple result forms the basis for
all subsequent analytical calculations. For example, it implies that the local
transverse mean-square displacement is given by $\lavg \vect r_\perp^2 \ravg =
L_d^3 / 4 \ell_p$.

In general, however, analytical calculations are not feasible. In order to
investigate the full range of parameters we employed a standard Monte Carlo (MC)
scheme using a discretized version of the wormlike chain model. This allows us
to go beyond the WBR limit and calculate a range of observables which are
directly accessible in single-molecule experiments. The polymer was represented
by a chain of $N$ segments $\vect t_i$ approximating the continuous contour. The
inextensibility constraint was imposed along the whole polymer by fixing the
segments length to the value $L/N$. The cylindrical symmetric harmonic potential
was calculated at the end points of each segment.

During the simulation, both ends of the polymer were assumed to be completely
free, in both position and orientation. The initial configuration was chosen
sufficiently close to full stretching, in order to ensure a fast convergence of
the MC algorithm. A new configuration was generated by changing the orientation
of a randomly chosen segment and accepted according to the standard Metropolis
algorithm. We have not considered effects resulting from self-avoidance, which
is not important for strong confinement and in the stiff limit, and is negligible
even in the flexible limit if the number of segments is below $N\approx500$. At
least $10^6$ MC steps per segment were performed, to obtain averages and
statistical errors. Our MC procedure was validated by evaluating known
quantities for polymers in bulk~\cite{FootnoteValidation}.

% fig_2
\begin{figure}[t]
  \centering
  \includegraphics{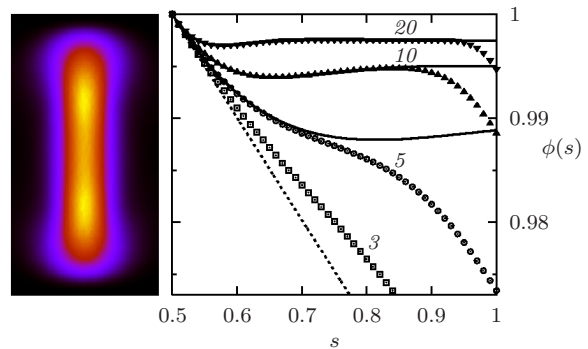}
  \caption{(Color online) \emph{Left:} marginal probability distribution function of polymer
    configurations in a plane containing the tube axis for $c=4$ and
    $\varepsilon =2$. The probability density increases from dark to light
    colors. \emph{Right:} normalized tangent-tangent correlation function
    $\phi(s)=\lavg\vect t(s)\cdot\vect t(0.5)\ravg$ for $\varepsilon=0.1$ and
    various confinements ranging from $c=3$ to $c=20$. MC simulations are
    represented by symbols, whereas the solid lines show the analytical
    approximations of $\phi(s)$ as from
    Eq.~\eqref{eq:ttcor_harmpot_weakly_bending}. Dashed line: exponential decay
    of an unconfined chain.}
  \label{fig:tt_corr}
\end{figure}
% fig_2

% Density profiles and tangent-tangent correlation function
In a typical experimental setup measuring the shape of a biopolymer in confined
geometry, the filaments are labeled with some fluorescent dye and recorded over
the time resolution window of the camera. This results in an intensity profile
for the emitted light that corresponds, in our theoretical model, to the
marginal probability distribution function of the positions of the constituent
segments in a plane containing the tube axis.  Figure~\ref{fig:tt_corr} shows this
function as obtained from our MC simulations for intermediate values of the
collision and stiffness parameters: $c=4$ and $\varepsilon = 2$. This picture
nicely illustrates the shape of a biopolymer in confined geometry. Given
sufficient experimental resolution it may even be possible to resolve the
bimodality of the distribution clearly visible in the MC data.

Beyond such a qualitative impression of the shape of the polymer, other more
quantitative measures may characterize better its conformations. We start our
discussion with the tangent-tangent correlation function $\phi (s) = \lavg
\vect t(s)\cdot\vect t(0.5) \ravg$, measured from the center of the filament.
Figure~\ref{fig:tt_corr} shows the results of our MC simulations in the stiff
regime for $\varepsilon = 0.1$ and a range of collision parameters $c$. For
small $c$ the data show the expected exponential decay of a free filament. For
$c \geq 1$ confinement effects become visible and deflections start to affect
the correlations for distances comparable to the Odijk length scale $1/c$. For
strong confinement, $\phi (s)$, after an initial exponential decay and a shallow
minimum at $s-0.5 \approx 1/c$, reaches a broad plateau before correlations
decay again in a small boundary layer of size $1/c$. All these features, but the
boundary layer effect, are well captured by a formula (solid line in
Fig.~\ref{fig:tt_corr}) easily obtained in the WBR approximation from
Eq.~\eqref{eq:mode_spectrum}:
\begin{multline}  \label{eq:ttcor_harmpot_weakly_bending}
  \lavg\vect t(s)\cdot\vect t(s')\ravg =
  1-\frac{\varepsilon}{2\,c}\left[\sqrt2\,\exp\left(-\frac{|s-s'|}{L}c\right)
  \right.\\
  \left.\times\sin\left(\frac{|s-s'|}{L}c-\frac\pi4\right)+1\right]\,.
\end{multline}

% fig_3
\begin{figure}[t]
  \centering
  \includegraphics{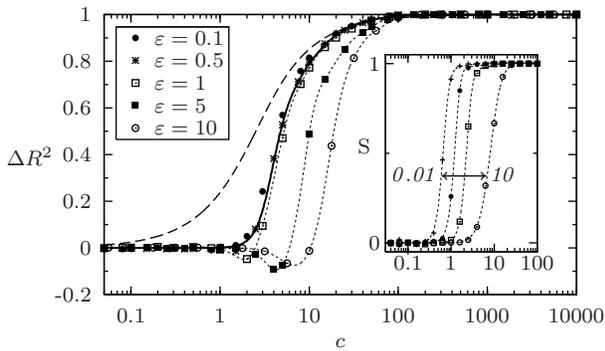}
  \caption{Reduced mean-square end-to-end distance $\Delta R^2$ as a
    function of the collision parameter $c$ for a series of flexibilities
    $\varepsilon$ indicated in the graph. The universal scaling curve (solid
    line) asymptotes the analytical result (long-dashed line) in the limit of
    strong confinement. Short-dashed lines are guides to the eye for
    $\varepsilon > 1$. Inset: orientational order parameter $S$ as a
    function of $c$ for $\varepsilon = 0.01, \, 0.1, \, 1, \, 10$.}
  \label{fig:scaleorient_prl}
\end{figure}
% fig_3

The quantity which best characterizes the elongation of the polymer is the
mean-square end-to-end distance $\lavg R^2 \ravg$. As already noted, this is
exactly known for unconfined chains, $ \lavg R^2 \ravg_0 = 2 L^2
\varepsilon^{-2} \left(\varepsilon - 1 +e^{-\varepsilon} \right)$, but can be
calculated for strong confinement in the WBR approximation. We find:
\begin{equation}
 \frac{\lavg R^2 \ravg_c}{L^2} = 1-\frac{\varepsilon}{2\, c} \left\{ 1 +
 \frac{1}{c^2}\left[1-\sqrt2\, e^{-c} \sin \left(c + \frac\pi4\right)\right]
 \right\} .
\end{equation}
The validity of this formula, obtained in the WBR limit, extends to 
the case of weak confinement: this result correctly recovers the exact 
result for unconfined chains in the stiff limit. Upon defining the 
reduced end-to-end
distance 
\begin{equation}
  \Delta R^2 (c) = \left(\lavg R^2\ravg_c - \lavg R^2\ravg_0\right) /
               \left(L^2 - \lavg R^2\ravg_0\right)\,,
\end{equation}
one finds that this quantity is a function of the collision parameter $c$ only.
This suggests to look for a data collapse in the MC data. In fact, as can be
inferred from Fig.~\ref{fig:scaleorient_prl}, the reduced end-to-end distance is
a function of the collision parameter only once the flexibility parameter falls
below $\varepsilon \approx 1$---i.e., in the stiff regime. This implies that
there is a single master curve characterizing the shape of a stiff polymer.

The analytical results capture the MC results only in the limit of very strong
confinement. This is due to the fact that the WBR approximation assumes the
filament to be perfectly aligned with the tube axis, which is strictly valid
only if $c \gg 1$.  In the regime of weak confinement the primary effect of the
geometric constraints is to align the filament with the tube axis. We expect
this alignment to start once the filament length (more precisely the end-to-end
distance) becomes comparable with the tube diameter---i.e., for $c^3 \sim
\varepsilon$. To render this statement quantitative we define the
orientational order parameter
\begin{equation}
  S = \frac12 \left( 3 \lavg \cos^2 \vartheta \ravg - 1 \right)\,,
\end{equation}
where $\vartheta $ is the angle of $\vect R$ with respect to the tube axis. In
fact, as can be inferred from the inset of Fig.~\ref{fig:scaleorient_prl}, there
is an intermediate confinement regime where the onset of orientational order
precedes filament elongation. This has important implications for the
confinement of biopolymers like F-actin and microtubules in artificial channels
and cellular systems. For instance, one estimates that for F-Actin with
$L=2 \mu$m this window in tube dimensions ranges from $d=2 \mu$m down to
$d\approx 0.4 \mu$m. In this window the free energy cost for confinement will
not be given by the Odijk estimate $F \sim k_B T c$ but by the constraint on the
orientational degrees of freedom---i.e., $F \sim k_B T \ln (L/d)$. This
intermediate regime becomes less pronounced with increasing flexibility.
Actually, even very long (self-avoiding) polymers are known to have an
instantaneous prolate shape~\cite{AronovitzNelson,Haber_Ruiz_Wirtz}. This
anisotropy in the radius of gyration tensor is rather due to entropy
effects~\cite{Kuhn} than the energy of bending as for stiff biopolymers
discussed here. Despite the different physical origin its consequences are that
also flexible polymers in confinement orient first before changing their
shape~\cite{VlietBrinke1990}.

For flexible polymers, with $\varepsilon \geq 1$, the initial effect of
confinement is to increase $R_\parallel$ but decrease $R$. There is a clear dip
in $\Delta R^2$, which can be explained by the following geometric argument.
Consider an initially randomly oriented end-to-end vector $\vect R$. Then weak
confinement will predominantly reduce the magnitude of the component of this
vector perpendicular to the tube axis but leave the parallel component
unchanged. This obviously leads to a decrease in the magnitude of the end-to-end
distance. We have checked this argument by measuring the projection of the
mean-square end-to-end distance onto the tube axis (data not shown), which
indeed does not show an initial decrease but increases monotonically. A dip in
the radius of gyration $R_G$ (not the end-to-end distance) has previously been
reported for very long ($\varepsilon \gg 1$) self-avoiding
polymers~\cite{VlietBrinke1990}. There it was argued to be a consequence of an
intermediate regime, where the principal components of the radius of gyration
tensor are reduced before confinement excluded-volume effects lead to an
increase in $R_G$.

% fig_4
\begin{figure}[t]
  \centering
  \includegraphics{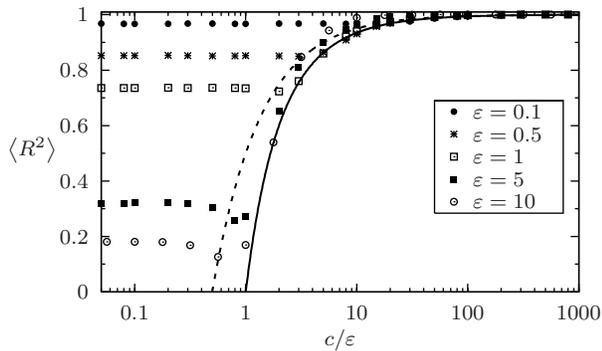}
  \caption{Scaling plot for the mean-square end-to-end distance 
    $\lavg R^2\ravg$ versus $c/\varepsilon = \ell_p / L_d$ for various
    flexibilities $\varepsilon$. Universal scaling curve as estimated from the
    simulation (solid line) and obtained in the WBR limit (dashed line).}
  \label{fig:scale_harm_prl}
\end{figure}
% fig_4

In the limit of strong confinement, where the Odijk deflection length is either
much smaller than the total length (for stiff polymers) or much smaller than the
persistence length (for flexible polymers), one expects $\lavg R^2 \ravg$ to
become independent of the filament length. In order to show this explicitly we
have replotted our MC data in Fig.~\ref{fig:scale_harm_prl} as a function of the
number of collisions within a persistence length, $c/\varepsilon = \ell_p /
L_d$. Indeed, all curves for different flexibilities merge into a master curve
for strong confinement. In contrast to the master curve for weak confinement,
which applies only for stiff chains, this master curve captures all chain
flexibilities. The numerical result asymptotes the analytical formula obtained
in the WBR limit, but only for $c/\varepsilon$ quite large. This shows that for
intermediate confinement the orientation and the less constrained ends of the
filament contribute significantly to the conformations; both of these effects
are neglected in the WBR limit.

Advances in microfabrication and nanofabrication technologies have made it possible to
confine biopolymers to topographical structures whose dimensions are
comparable to or even smaller than their persistence length. This opens a range
of novel possibilities to visualize and manipulate DNA and cytoskeletal
filaments. Here, we have presented an extensive theoretical analysis of the
shape and conformations of biopolymers resulting from strong confinement and
identified and quantified a range of novel scaling regimes. We make specific
predictions for experimentally accessible quantities like the density profile or
the orientation and apparent length of a biopolymer in a channel. Our
calculations may provide a road map for a clear identification of the possible
scaling scenarios involved in the manufacture of nanofluidic and microfluidic
devices.  At the same time, our analysis is a first step towards a quantitative
master curve connecting the apparent length and the actual length of DNA in
nanochannels, which has important implications for experimental realizations
aimed at a rapid screening of entire genomes in contrast to gel electrophoresis.
Finally, they shed some light on the effect of cellular crowding on the
conformation of cytoskeletal filaments.

\vspace{2ex}
We thank Tobias Munk for helpful discussions. We acknowledge financial support
from the Marie Curie contract MERG-CT-2004-513598 and PRIN 2005 (GL) and the DFG
through grant SFB 486 (EF). Financial support of the German Excellence
Initiative via the program ``Nanosystems Initiative Munich (NIM)'' is gratefully
acknowledged.

\end{document}